\documentclass[12pt,a4paper]{article}
\usepackage{amsmath,amssymb}
\usepackage{graphicx}
\usepackage{cite}

\allowdisplaybreaks[3]

\begin{document}
\begin{flushright}
KEK-TH-2613
\end{flushright}

\begin{center}
{\Large\textbf{
Fokker-Planck Equation 
and  de Sitter Duality}}
\end{center}

\begin{center}
{Yoshihisa\textsc{Kitazawa}}
\footnote{E-mail address: kitazawa@post.kek.jp} 
\end{center}

\begin{center}
\textit{KEK Theory Center, Tsukuba, Ibaraki 305-0801, Japan}\\
\textit{Department of Particle and Nuclear Physics}\\
\textit{The Graduate University for Advanced Studies (Sokendai)}\\
\textit{Tsukuba, Ibaraki 305-0801, Japan}
\end{center}

\begin{abstract}
 Infra-Red scaling property of inflationary universe is in the same universality class of random walk.
 The two point correlators of  the curvature perturbations are enhanced by the e-folding number $N$.
The distribution function of the curvature perturbation $\rho_t (\zeta)$ satisfies the Fokker Planck equation. The de Sitter universes are dual to the random walk: They belong to  the  Universality class of  dimension two fractal.
These boundary and bulk duality are  at the heart of holography 
of quantum gravity. 
Historically the  correspondence of thermodynamics and Einstein's equation are recognized as the first evidence for de Sitter duality .
Our de Sitter duality relates the stochastic and geometric point of view.
We study two types of the  solutions of FP equation in quasi de Sitter space: (1) UV complete spacetime and (2) inflationary spacetime with concave potentials. 
The maximum entropy principle favors the following scenario: The universe is (a) born with small $\epsilon$ and (b) grows by inflation in the concave potential. 
We predict $n_s\leq 0.975(0.97) $ and $r \leq 0.04(0.03)$ for $N=50(60)$ at the pivot angle $0.002Mpc^{-1}$. 
We have lowered  the upper bound of $r$  by taking account of random walk effect at the  boundary.
Our predictions are highly consistent with recent observations. 
\end{abstract}

\newpage

\section{Introduction}
\setcounter{equation}{0}

In four dimensional de Sitter space , duality plays an important role.
We study the Fokker Planck equation 
on the boundary and its implication on the bulk dynamics  and geometry .
We study stochastic  and geometric duality in de Sitter space.

We have adopted BRST type local gauge.
In our parametrization and gauge, the massless scalar field with the canonical kinetic term is
\begin{align}
X=(2\sqrt{3})\omega-(\frac{1}{\sqrt{3}})h^{00}
\end{align}
We may focus on the  massless minimally coupled $X$ field to explore IR dynamics.

The curvature perturbation $\zeta$ 
is the fluctuation of the over all scale of the spacial metric:
 $g_{ij}=exp(2\zeta\delta_{ij})(\hat{g})_{ij}$.  
 Indeed we find $X$ is equal to curvature perturbation $\zeta$.
\begin{align}
\zeta= \sqrt{3}h^{00}-\frac{1}{\sqrt{3}}h^{00}=X.
\end{align}
The conformal mode contains $X$ field with the following ratio.
\begin{align}
2\omega=h^{00}= \sqrt{3}/2X
\end{align}

@`See Appendix A for more details on our conventions.  

The Einstein gravity consists of massless minimally coupled modes 
and conformally coupled modes. 
We recall de Witt-Schwinger expansion of a scalar propagator in de
Sitter space.
\begin{align}
<\zeta\zeta>=&-(G_0a_0 + G_1a_1 +\bar{G} )(16\pi G)
\end{align}

Although the leading term is  quadratically divergent,
 it can be subtracted in the dimensional regularization. 
We find infrared logarithmic singularity when $a_1=R(1/6-\xi) $ does not vanish. $\xi$ is a tunable parameter. We are most interested in the minimal coupling case $\xi=0$. The conformal coupling corresponds to $\xi=1/6$. 

\begin{align}
<\zeta^2>=&
- \frac{G_N R}{ 6\pi}\frac{k^{D-4}}{4-D}\\
  =&-4g \frac{1}{(4-D)}+ 4g \log k
 \end{align}
 It is logarithmically divergent near 4  dimensions as $1/(4-D)$ pole indicates .

 The renormalization clarifies  
 $ <\zeta^2>_R$  as $4gN^*$ where $N$ is the e-folding number.
 This $N$ dependence of  the coming back probability $<\zeta^2>$ is in agreement with the Langevin equation  in the next section.
As we emphasize, there is no time dependent UV contributions. 
We have focused on the Hubble scale physics where 
$He^N=Ha=k$.
Our renormalization scale is
$\log k^*\sim N^*$. 
The star implies the quantity is evaluated 
at the time of horizon crossing $N^*$.

In the conformal coupling case, the second coefficient of the de Witt-Schwinger expansion of the propagator cancels in de Sitter space
as $a_1 =R/6-2H^2=0$. We find no IR logs in this case.
We  focus on the minimally coupled modes where
$a_1=R/6$. 

We diagonalize the conformal mode $\omega$ and $h^{00}$
as follows. 
\begin{align}
X\equiv 2\sqrt{3}\omega
-\frac{1}{\sqrt{3}}h^{00},\hspace{1em}
Y\equiv h^{00}-2\omega. 
\end{align}
$X$ is the massless minimally coupled modes.
It turns out to be nothing but the curvature perturbation $X=\zeta$.
 
We have computed the quantum correction to Einstein gravity at super horizon scale.
\cite{Kitamoto2019-2}.
{}
The  Einstein gravity belongs to the 
inflation theory since an
inflaton field  is indispensable to renormalize it.
Einstein gravity is dual to inflation theory in this sense. 
In section 2, large $N$ power counting rule  is introduced. 
The duality holds between  the random walk in the boundary
and slow roll inflation theory in the bulk.
The fractal dimension of random walk is 2 and universal.
This fact implies the two point functions
of massless scalar are enhanced by $O(N)$.
$<\zeta^2>\sim 4gN$.
 
 \section{Large N Power Counting}
\setcounter{equation}{0}

The gravity has always been investigated from geometric point of view following Einstein. Inflation theory is no exception.
On the other hand, the inflation theory leaves too many unexplained freedom. What is the inflaton? Why they perform slow roll on the potential? What determines the potential? and so forth.

There is an alternative (dual) approach, i.e. stochastic point of view.
 The  Fokker-Planck equations  on the boundary play the dual role to the Einstein's field equation.
 Thermodynamics provide one of Einstein equation ${1/2}\dot{f}^2=\dot{H}^2$.
The FP equation gives the identical solution.
This duality may be capable to explain many puzzles we have raised in the beginning of this section.  Namely slow roll of inflaton occurs by the collisions with   the curvature perturbation.
This idea enables us to predict the large $N$ power counting  of the slow roll parameter: $\epsilon \sim 1/N$ . 

Suppose we consider a concave potential, 
\begin{align}
& g\propto N^{1/2n}\propto f^{n}, >1,\\
2\epsilon =&-\frac{\partial}{ \partial N}\log g=1/2n \\
\epsilon &<1/4N .\end{align}

The Brownian system consists of the macroscopic particles and numerous atoms of water .
The analogy to draw here is {} 
CMB temperature distribution  of local patches and the curvature perturbation respectively. The former is labeled by the small temperature fluctuations: 
$\delta T/T \sim 10^{-5}$. $g$ is also small $10^{-10}$.
We recall the Langevin formalism next a la
\cite{Parisi}.

\begin{align}
\dot{\zeta}=b(t).
\end{align}
In Einstein gravity, there is only one massless scalar mode $X$ as explained in the introduction.
\footnote{The conformal mode is the mixture  of massless ($X$) and massive ($Y$) modes. The cosmological constant and Scalar curvature operators are dressed by the conformal mode. The correlator $<X^2>$  renormalize the cosmological constant. See Appendix A.}   

Let us consider 
\begin{align}
B(t)^{\delta}=\int_t^{t+\delta}dt' b(t').
\end{align}
We assume $\delta$ can be taken arbitrary small.  
In the low energy approximation, we can ignore the drift term.
This assumption also holds as there is no drift term in $\zeta$ direction.

Under such circumstances, 
\begin{align} 
\overline{b(t_1)b(t_2)}=\delta(t_1-t_2)4Hg
\label{Lancor}
\end{align}

The scalar variance is estimated in the  Langevin formalism.
\begin{align}
\overline{(\zeta(t)-\zeta(0))^2}
=&\int_0^tdt_1\int_0^tdt_2\overline{b(t_1)b(t_2)}\\
=&\int_0^t dN 4g\sim 4gN
\label{SV}
\end{align}

In the differentiated form: 
\begin{align}
\frac{\partial}{\partial N}\overline{(\zeta(t)-\zeta(0))^2}=4g(t)
\end{align}

The important point is that only massless scalar variance is enhanced by $N$ through the Brownian motion.
There is no such effects other than the scalar variance due to Lorentz symmetry.
The non-singlets such as gravitons carry helicity $\pm 2$. 
In contrast, the Lorentz symmetry is gauged in quantum gravity and it protects non-singlet states.  The elementary particles and the vacuum are the representations of the Poincare group .
For example the tensor contribution is simply given by  their tree  contribution.

 Tensor  to  Scalar ratio is  estimated as follows.
 We first assume $g$ is slow and small to be treated as a constant.
\begin{align}
\frac{T}{S}=
\frac{8g}{4N(t)g} 
= \frac{2}{N(t)}.
\label{BRST}
\end{align}

The numerator in (\ref{BRST}) can be calculated by perturbation theory .
The denominator in (\ref{BRST}) is  given by Langevin equation .
It is because macroscopic scalars' are labeled by the additional quantum numbers, namely, spins.

$\tilde{N}=N_e-N$ where $N_e$ denotes the e-foldings when the inflation terminates. In order 
to keep inflation going, we may assume $1<\tilde{N}<60$.

\footnote{This is the upper limit of 
$T/S$ ratio in the linear potential. It is the boundary between the convex and concave potentials. Such a point may be chosen naturally since the entropy is maximum. }

The  slow roll parameters  for the power potential models will be
worked out in more detailed in the next section.
We emphasize that the Brownian motion at the horizon is the sole
source of anisotropy of CMB. Inflaton  in bulk gravity is de Sitter dual to Brownian motion on the boundary.

The 2 point function $<\zeta^2(t)>$ is
 the diffusion kernel if
$g(t)$ changes much slower
  than the distribution function. It is well approximated to be  Gaussian with the standard deviation $4gN(t)$. The slow roll condition is satisfied .
However we shall construct cosmologically more realistic  solutions in the subsequent sections.  They include inflationary universe with the concave potential and $UV$ finite space-time . These models are $t$ dependent  as they contain time dependent gravitational coupling $g(t)$ .
 
 Remarkably  the concave potential is the only possible stage for the inflation.
The boundary theories are the holographic solutions of  FP equations in $1+1$ dimensions. de Sitter duality implies the 4 dimensional slow roll inflationary
universe  is dual  to 2 dimensional random walk. 
Our formulation of de Sitter duality may leads to Holographic fluctuation- dissipation theorem  .  It is because fluctuations of the metric screens the gravitational coupling $g(t)$. If $g(t)$ decreases, the
entropy  inevitably increases : $S=1/g(t)$. As de Sitter entropy is the ratio $g\sim H^2/M_P^2$,  the entropy of the Universe is just enormous $10^{120}$. 

So far, we observe large $N$ enhancement of some correlators in de Sitter space.
They arise as IR logarithms in stochastic theory. 
The resummation of IR logarithms by the renormalization group  is carried out  by the FP equation.
 The 2 dimensional boundary theory projects the hologram from the history of entropy and inflation  in particular. The space-time structure is determined
 by the time dependence of $g(t)$.
In de Sitter space, the infra-red quantum 
fluctuations may give rise to time dependent effects such as shielding the cosmological constant.

 The Einstein's equation follows from the first law.
Historically the correspondence between thermodynamics and Einstein's equation is the first evidence for de Sitter duality   \cite{Jacobson}.
Obviously we are estimating the entropy of sub-horizon $S_A(g(t))$.
de Sitter temperature $T$, Hubble parameter $H$ are related as 
 $T=1/R,H=1/R,T=H/2\pi$ . \cite{2018}.

The renormalization group is one of the most powerful tool to investigate the
resummation of IR log type contributions. 
The important point is that there is only one massless scalar mode  which
plays the significant role as explained in the introduction. 
We call it  $X$ field although it is nothing but the curvature perturbation $\zeta$.

The remarkable relation $S_A=S_B$ holds when we consider the 
$A,B$  
as the boundary-bulk pair $A=\partial B$. 
The entropy of bulk (sub-horizon) and boundary (super-horizon) are the same.
This UV-IR symmetry might explain that the FP equation posesses UV finite solutions.
It further indicates 
the Universe starts with the de Sitter expansion near the Planck scale with 
$\beta\sim\epsilon=0$. 

In order to elucidate the IR logarithmic effects non-perturbatively, 
we formulate the Fokker Planck equation for the curvature perturbation of the metric. 
We obtain the $\beta$ function of dimension-less gravitational
coupling $g=G_NH^2/\pi$,
 $G_N$ is the Newton's coupling and $H$ is the Hubble's  parameter. 

We first identify the
one-loop running coupling:  $g=2/(\log N) $. It  logarithmically decreases  toward the future in the leading $\log$ approximation. 
We then consider the sub-leading corrections which are suppressed by 
$g$.

Since the $\beta$ function turns out to be negative, $g$ is 
asymptotic free toward the future
\cite{Gross1973,Politzer1973}. 
Furthermore, the $\beta$ function possesses the ultraviolet (UV) fixed point with the critical coupling $g=1/2$. 
The theory in tern satisfies  asymptotic safety  around the Planck scale.\cite{Weinberg}.

This scenario indicates that our Universe begun with the dS expansion  at the Planck scale with a minimal entropy (1-bit).

Since $\epsilon\propto \beta$ holds, the  smallness of
$\epsilon$ in the beginning of the Universe is naturally guaranteed 
in our scenario. We have postulated
the quantum correction to de Sitter entropy is given by the Gibbs entropy of the curvature perturbation.
We may decompose the total Hilbert space  $H=H_A \otimes H_B$ into   $H_A$ and $H_B$. 
Quantum definition of $\rho_B$ involves the integration on A: 
$ \rho_B=Tr_A\rho^{tot}$.
This operation corresponds to the renormalization of $\rho$ at the horizon scale.
Then 
$S_B=-Tr_B\rho_B\log(\rho_B)$.
This formula counts the entropy of the massless sector
 in the super-horizon.  
  At the quantum level, the de Sitter entropy is von Neumann entropy, i.e.  entangled entropy: (EE).
  It equals the entropy of sub-horizon $S_A=S_B$ C for more information.

\section{de Sitter Entropy and Curvature Perturbation}
\setcounter{equation}{0}

The infrared logarithms are important clue to elucidate de Sitter entropy.
The FP equation is very effective to resum IR logarithms. 
The solutions of FP  equation consist of  the time dependent distributions for $\zeta$.
 We find the standard deviation $\sigma^2=<\zeta^2>$ possesses the  time dependence $g(t)N(t)$  with the preceding Langevin formalism. 
Our proposal is the CMB distribution function evolves just  like that of the macroscopic particles in the Brownian motion. Since the fractal dimension of random walk is $2$,
the equal time  2 point function $<\zeta^2>$ scales as $O(N) $.
We first work out the leading log solution which 
 is valid in the large $N$ limit. As long as the gravitational coupling $g$ is constant,
the FP equation  shows that the solution obeys the diffusion equation with the  standard deviation  enhanced by $O(N(t))$ as has been shown in section 2 .

 The Entangled Entropy of the curvature perturbation  increases logarithmically
 \cite{Kitamoto2019-2} .
\begin{align}
-Tr {\rho\log \rho}
\sim\frac{1}{2}\log N(t)
\label{entropy}\end{align} 
We assume $g(t)$ changes much slower than $N$ 
which turns out to be self-consistent  (\ref{betag}).

Identifying the EE of curvature perturbation
with the quantum correction to dS entropy,
we obtain the bare entropy $S_B$ with the counter term
to cancel the quantum correction.
Since $S(t)=1/g(t)$ where $g(t)$ is the dimension less gravitational coupling,
\begin{align}
\frac{1}{g_B}=\frac{1}{g(N)} - \frac{1}{2}\log(N). 
\label{coupling}\end{align}
By requiring the bare coupling is  
independent of the renormalization scale: namely  $N$ , as we have just 
done to derive FP equation,
we obtain the one loop $\beta$ function.

\begin{align}
\beta=\frac{\partial}{\partial \log(N)}g(N)=-\frac{1}{2}g(N)^2. 
\label{beta1}\end{align}
We find the running gravitational coupling in the leading log approximation:
\begin{align}
 g(N)=\frac{2}{\log(N)}, 
\label{betag}
\end{align}
It agrees with de Sitter entropy \cite{Gibbons1977}.
 \begin{align}
 S=1/g(N).
 \end{align}
 The list of the related semi-classical relations is quite extensive.
 
Area of horizon:$A=4\pi/H^2$.
entropy: $S=A/4G=\pi/GH^2=1/g$.
At present: $GH^2=10^{-60}\times 10^{-60} \sim 10^{-120}$.

 We regard these remarkable  relations as the strong evidences that de Sitter entropy is EE of curvature perturbation.
  In order to determine the time evolution of entropy,
  we need to use time independent UV cut-off like dimensional regularization to identify time dependent quantum corrections. 

We have evaluated the time evolution of EE to the 
leading log order in \cite{Kitamoto2019-2}.
We need to be careful here to correctly estimate higher loop corrections.
For that purpose, the FP equation 
should be renormalized to all orders.
This is one of our main achievements on the identity of de Sitter entropy. 

The bare distribution function is constructed by subtracting the bulk mode contributions . It is independent from 
the renormalization scale or exit time .
The subtraction of UV divergences from $\rho$ is accomplished to all orders by multiplying the following operator from the left: 
\footnote{The proof of this statement is given in \cite{Holbeta}.}

\begin{align}
\rho_B=\exp\Big(2g(t)(\frac{k^{D-4}}{4-D})\frac{\partial^2}{\partial\zeta^2}\Big)\times\rho .
\label{bare}\end{align}
Here $k$ is the renormalization sale, $N=\log k$.
As $\rho_B$ is independent of the exiting time, ${\partial\frac \partial N}{\rho_B}=0$,
the distribution function $\rho$ obeys the following renormalization group equation.
\begin{align}
{\partial\frac \partial N}\rho_N(\zeta)
=2g\frac{\partial^2}{\partial \zeta^2}\rho_N(\zeta)
\label{FP0}\end{align}.

We have derived the  FP equation as the renormalization group equation.
FP equation is consistent with Langevin equation. 
\begin{align}
{\partial\frac \partial N} <\rho_N(\zeta) \zeta^2>
=  2g\frac{\partial^2}{\partial \zeta^2}<\rho_N(\zeta) \zeta^2>
=4g
\end{align}

  The distribution function which is the  solution of the equation (\ref{FP0}) is 
\begin{align}
\rho_N=
\frac{1}{\sqrt{8\pi g(t) N(t)}}\exp({-\frac{\zeta^2}{8g(t)N(t)}})
\label{difsol}\end{align}
This is a Gaussian distribution with the standard deviation $4g(t)N(t)$.
The principal goal of this paper is  to extract physical  implications from de Sitter duality, the consistency of stochastic and geometric descriptions. 
The above distribution function is valid when $g$ changes much slowly than $N$ in agreement with
(\ref{betag}).
Gibbons-Hawking de Sitter entropy is understood as as the semi-classical 
solution of FP equation.

In the inflationary universe,
the energy flux through the horizon is given by the  integral  
 \begin{align}
 \delta E = \delta gd\Sigma_{\mu}  T^\mu_\nu\xi^\nu , 
  \end{align}
 where $d\Sigma\mu$ is the 3-volume of the horizon, and $\xi_\nu$ is the null generator of the horizon.

We will equate the energy flux through the horizon to the change of geometrical entropy dS.
In the dual inflation theory picture, de Sitter entropy increases due to the incoming energy flux of inflaton. 
The speed of increase of the entropy is estimated
by the first law $T\Delta S=\Delta E$ where $\Delta E$ is the incoming energy flux of inflaton. 
After translating the change of entropy into that of the Hubble parameter by Gibbons-Hawking formula, 
one of the Einstein equation is obtained \cite{Frolov2003}, 

\begin{align}
\ddot{\rho}=-\frac{1}{2}\dot{f}^2. 
\label{hot}
\end{align}

In the Einstein gravity picture, the reduction of Hubble parameter and entropy creation takes place 
due to incoming inflaton energy flux. 

This classical picture is dual to the stochastic picture: 
quantum diffusion of the Hubble parameter and stochastic creation of entropy. 

 (\ref{hot}) can be re-expressed in terms of the slow roll parameters: 
\begin{align}
2\epsilon
=\frac{\partial }{ \partial \tilde{N}}\log g(t).
\label{GFP}
\end{align}
Since (\ref{hot}) and (\ref{GFP})
reproduces the Friedman's equation of motion in the inflation theory, the solution must respect the general covariance.

We identify two different groups of the solutions:  1. UV complete solutions and 2. inflationary solutions .
 FP equation on the boundary and slow roll inflation  in the bulk
constitute de Sitter duality. 
Our boundary theory is identical to that of the Brownian motion.
It is holographic description of  $dS_4$ gravity. In our theory, the
curvature perturbations trigger random walk on the boundary (horizon).
 $g$ scales as
$g\propto \exp(-2\epsilon N) \sim a^{-2\epsilon}$ 
where $a=e^{N}$ due to IR logarithmic effect at the one loop level.
 \cite{Kitamoto2019-2}  \cite{BCFH}.
 Since  the inflaton rolls  down  the potential $g\sim a^{-2\epsilon}$ generated
 by the quantum loop effects, it is  legitimate to claim the slow roll of inflaton  is
quantum effect. See Appendix A for a review on this process.

The tilt of the gravitational wave spectrum is
given as (\ref{GFP}).

We obtain the following anomalous dimension for a scalar field from the Langevin expression (\ref{SV}).
\begin{align}
\frac{\partial}{\partial \tilde{N}}\log(4g \tilde{N})
=  2\epsilon+\frac{1}{\tilde{N}}
\end{align}

It is because  the scalar field performs the Brownian motion and Langevin description takes care of this effect accurately for the fixed world line length 
$\tilde{N}=1/(4\xi)$.
Using this relation, we conclude the scalar fields
 of inflationary universe satisfy
 the following FP relation 
\begin{align}
6\xi=\frac{\partial}{ \partial \tilde{N}}\log \frac{g}{\xi}
= 2\epsilon+\frac{1}{\tilde{N}}
\label{FPIn}
\end{align}.

The solutions with power potentials are considered in the next section from  de Sitter duality point of view.
Our basic conjecture is that the  slow roll of inflaton is  caused  by the collisions  
with the curvature perturbation.
This idea explains generic features and may allow deeper penetration into  de Sitter duality.

The attractive point of our theory is its simplicity. 
The inflationary universes diffuse slowly due to the random collisions with the curvature perturbation. The universal enhancement of scalar to tensor ratio in the large $N$ limit is explained by its fractal dimension of the scalar modes. 
The  $2$ point function goes like 
\begin{align}<(\zeta_i)^2>\sim  4g N
\end{align}
--
 
 There is no UV fixed point at the leading log order.
However 
there is a UV fixed point in our renormalization group at the 2-loop level.
FP equation enables us to evaluate higher order corrections to the $\beta$ function. 
We resum IR logs  to all orders. 
We can confirm that the following $g_f$ and $\xi_f$ satisfies (\ref{FPIn}), 
\begin{align}
g_f=&\frac{2}{\log N}\big(1-\frac{1}{\log N}\big),\\
{\xi}_f=&\frac{1}{6N}\big(1-\frac{1}{\log N}\big). 
\label{solution1}\end{align}

The one loop running coupling  (\ref{betag}) is justified
 in the large N limit.
\begin{align}
g=\frac{2}{\log N},~~
{\xi}=\frac{1}{6N}. 
\end{align}

Thus, the $\beta$ function, $\epsilon$ and the semi-classical entropy $S$ are given by 

\begin{align}
\beta=\frac{\partial g}{\partial \log N}
=-\frac{2}{\log(N)^2}+\frac{4}{\log(N)^3}.
\label{beta2}\end{align}

\begin{align}
\epsilon =& -({1/ 2}){\partial \log(g)/ \partial N}\\
=&-\frac{1}{2gN}\beta \sim\frac{1}{2}{N\log N}
 \end{align} 
 
\begin{align}
\frac{\partial}{\partial N}S_{cc}=
\frac{\partial}{\partial N}\frac{1}{g_f}
=-\frac{1}{Ng_f^2}\beta_f.
\label{entropy2}\end{align}'

A remarkable feature is that the coupling has the maximum value $g=1/2$ at the beginning. 
It steadily decreases toward the future 
as the $\beta$ function is negative in the whole region of time flow. 
On the other hand, $\epsilon$ increases toward future until inflation terminates at $\epsilon\sim O(1)$.
 At the UV fixed point,
$\epsilon and
\beta $ vanish at the beginning . 
The existence of the UV fixed point indicates the consistency of quantum gravity. 
The single stone solves the
$\epsilon$ problem \cite{Penrose1988} as well since $\epsilon$  vanishes at the fixed point $\beta=0$.

The $\beta$ function describes a scenario that 
our Universe started the dS expansion with a minimal entropy $S=2$.
while it has $S=10^{120}$ now. 

 Our results on the UV  fixed points are not water tight
as the coupling is not weak. Nevertheless we find it remarkable that they
support the idea that quantum gravity has a UV fixed point 
with a finite coupling. In fact 4 dimensional de Sitter space is constructed in the target space 
at the UV fixed point of $2+\epsilon$ dimensional quantum gravity  \cite{Kawai1993}.
4 dimensional de Sitter space also appears at the UV fixed point of the exact
renormalization group \cite{Reuter}\cite{Souma}.

Such a theory might be a strongly interacting conformal field theory .
However, it is not an ordinary field theory as the Hubble scale is Planck scale. 
Our dynamical $\beta$ function is closely related to the cosmological horizon and physics around it. 
The existence of the UV fixed point could solve the trans-Planckian physics problem. 
A consistent quantum gravity theory can be constructed 
under the assumption that there are no degrees of freedom at trans-Planckian physics \cite{Bedroya2019}. 
In this sense, it is consistent with string theory and matrix models.  
The Universe might be gfracned by (\ref{solution1}) in the beginning 
as it might be indispensable to construct the UV finite solutions of the FP equation.

\section{Concave Mode and Inflaton}
\setcounter{equation}{0}

The  distribution of the curvature perturbation is characterized by the standard deviation $1/\xi$ . Although there is no inflaton in Einstein gravity, 
we propose to identify the inflaton $\varphi^2$ as $ 1/\xi$. 
In our interpretation, the inflaton is not a fundamental field but a soft scalar mode of the metric. 
It grows due to the IR logarithmic fluctuations $1/\xi\sim N $.
While the inflation theory is specified by the inflaton potential, 
the dynamics of quantum gravity is determined by the FP equation 
which describes the stochastic process at the horizon. 
We thus argue 
the classical solution of the inflation theory satisfies the FP equation as well. 

\begin{align}
\int \sqrt{-g}d^4x\frac {1}{16\pi G_N}
\Big[R-H^2V(\varphi)-\frac{1}{2} \partial_\mu \varphi\partial^\mu \varphi\Big]
\label{action2}\end{align}

Let examine the linear potential
$V(\varphi) = 1+\sqrt{\gamma}\kappa \varphi$.
At the tree level, the slow-roll parameters are 
\begin{align}
\epsilon=(V'/V)^2/(16\pi G_N)=\gamma
\label{slow}
\end{align}
where $\kappa^2=16\pi G_N, M_P^2=1/8\pi G_N$.
\begin{align}
2\epsilon=-{\partial\over \partial \tilde{N}\log V
\end{align}
It coincides with Friedman equation (\ref{GFP}). 
the classical  inflation theory (\ref{action2}) is dual to the  Einstein gravity as they may consistently account for the effects of the leading IR logarithms. 

We confirm the equality between the slow-roll parameter $\epsilon$ and 
the slope of the linear potential $\gamma$.
The Hubble parameter $H^2(t)/H^2$ behaves as 
\begin{align}
V(\varphi)=1+\sqrt{\gamma}\kappa \varphi=1-2\gamma Ht=1-3gHt. 
\label{tile}\end{align}
This should be compared with the scaling inflaton potential.
\begin{align}
V(\varphi)=\exp(\sqrt{\gamma}\kappa \varphi)\sim 1-2\gamma Ht. 
\end{align}  
The above coincidence  implies that 
we have successfully identified the whole leading IR logarithms .
Thus, the slow roll inflation theory  contains the quantized Einstein gravity
with the concave potential.
This fact constitutes a strong evidence for  dS duality 
between stochastic and inflation theory (or a quintessence theory). 

We briefly review Entangled entropy in Appendix B.
As the potential of inflaton must be  concave
\begin{align}
	V(N)= N^{1/2n}=f^{1/n} , 
\end{align}
\begin{align}
V(f)''=\frac{1}{n}{(\frac{1}{n}-1)}f^{\frac{1}{n}-2} < 0.
 \end{align}
 where $n>1$ is assumed. 
 Since
 $\frac{\partial }{ \partial \tilde{N}}\log {g}=1/2nN $,
 the power potential inflationary universe  (\ref{solution2})
 is the solution of GFP below.

\begin{align}
6\xi=\frac{\partial }{ \partial {N}}\log (\frac{g}{\xi})
\end{align}.
\begin{align}
g=\tilde{N}^\frac{1}{2n},\hspace{1em}, 
\label{solution2}\end{align}

In this form, it is evident that $6\xi$ is the scalar spectral index in agreement with slow roll inflation theory. 
It is $O(1/N)$ as the scalar 2 point function is enhanced by $O(N)$ relative
to tensor  2 point function. 
We have shown GFP can be derived by macroscopic arguments also.
From quantum gravity point of view, this is an important non-perturbative evidence for 
de Sitter duality: quantum/geometric duality

We adopt the same scale $0.002Mpc^{-1}$ for $r,n_s$ to respect self-consistency.
It is important to establish the bound on $n$.
We recall here the curvature perturbation:
\begin{align}
P\sim 2.2 \times 10^{-9}.
\label{CP}
\end{align}
It is bounded from below $g>10^{-11}$ if $\epsilon>1/200$.

\section{Cosmological Solutions}
\setcounter{equation}{0}

The von Neumann entropy of curvature perturbation  provides the quantum correction to the semiclassical dS entropy $S=1/g(t)$. 
 $g(t)$ undergoes time evolution due to the quantum effect,
i.e., a quantum inflation picture. 

We have investigated  two types of the solutions of GFP equation. 
The first category consists of the pre-inflation solutions (\ref{solution1}) which are UV complete .
The second category is the  inflationary space-time solution (\ref{solution2}) which is not UV complete ,
but the inflation terminates when $\epsilon$ grows up to $0(1)$. 
This transition scenario is consistent with the physical expectation that the solution with dominant entropy is chosen. 
Since the entropy $1/g$ for (\ref{solution1}) increases logarithmically 
and that for (\ref{solution2}) increases in a power-law, it is possible
the former is chosen initially and the latter is chosen if we wait long enough. 

The gravitational coupling may start with $g=1/2$ at the UV fixed point. 
The Universe may enter the inflation era. $50 < N <60, H/H_0 \sim 10^{-5}$ for linear potential.

Although it is an attractive scenario consistent with maximum entropy principle, it is another matter to reveal the transition 
process from pre-inflation to inflation theory.   

We need to clarify wether complete solutions exist or not. Another question is which combination of the solutions account for the Universe.
For UV completeness,  the pre-inflation solution (\ref{solution1}) is indispensable.
From the entropy considerations, the universe with small $n$ may be favored.
Although there are many interesting issues to be clarified, they are beyond the scope of this paper. 
Holography is a specific feature of quantum gravity.
In this paper we have derived the Fokker-Planck equation
on the boundary by using the technique of the renormalization group . We have proven the de Sitter duality between 
the bulk and the boundary located at the horizon. We find
de Sitter duality is deeply connected with the von Neumann / entangled entropy.
Suppose the total Hilbert space is factorized into
the bulk states and boundary states. The entropy of the bulk and
the boundary are the same. This explains why de Sitter duality exists at all.
It also explains why $\beta$ function at the UV fixed point
can be calculable from the FP equation at the boundary. 
Nevertheless the dual pair have the identical entropy .

It is also a very important clue that the Inflaton potential is concave
due to strong subadditivity of EE.
In fact, convex  potentials are observationally almost excluded by now.

The inflation is estimated to last                                                                                                                                                                                                                                                                                                                                                                                                                                                                                                                                                                                                                                                                                                                                                                                                                                                                                                                                                                                                                                                                                         $ 50 <N< 60$. 
On the other hand, $g\sim 10^{-10} $.
So we can regard the effect of $g$ to be tiny and safely negligible.
In the weak coupling regime $g<\epsilon$, the theory becomes classical.

The von Neumann entropy of curvature perturbation  provides the quantum correction to the semiclassical dS entropy $S=1/g(t)$. 
 $g(t)$ undergoes time evolution due to the quantum effect,
i.e., a quantum inflation picture. 

We have investigated  two types of the solutions of GFP equation. 
The first category consists of the pre-inflation solutions (\ref{solution1}) which are UV complete .
The second category is the  inflationary space-time solution (\ref{solution2}) which is not UV complete ,
but the inflation terminates when $\epsilon$ grows up to $0(1)$. 
This transition scenario is consistent with the physical expectation that the solution with dominant entropy is chosen. 
Since the entropy $1/g$ for (\ref{solution1}) increases logarithmically 
and that for (\ref{solution2}) increases in a power-law, it is possible
the former is chosen initially and the latter is chosen if we wait long enough. 

The gravitational coupling may start with $g=1/2$ at the UV fixed point. 
The Universe may enter the inflation era. $50 < N <60, H/H_0 \sim 10^{-5}$ for linear potential.

Although it is an attractive scenario consistent with maximum entropy principle, it is another matter to reveal the transition 
process from pre-inflation to inflation theory.   

We need to clarify wether complete solutions exist or not. Another question is which combination of the solutions account for the Universe.
For UV completeness,  the pre-inflation solution (\ref{solution1}) is indispensable.
From the entropy considerations, the universe with small $n$ may be favored.
Although there are many interesting issues to be clarified, they are beyond the scope of this paper. 
Holography is a specific feature of quantum gravity.
In this paper we have derived the Fokker-Planck equation
on the boundary by using the technique of the renormalization group . We have proven the de Sitter duality between 
the bulk and the boundary located at the horizon. We find
de Sitter duality is deeply connected with the von Neumann / entangled entropy.
Suppose the total Hilbert space is factorized into
the bulk states and boundary states. The entropy of the bulk and
the boundary are the same. This explains why de Sitter duality exists at all.
It also explains why $\beta$ function at the UV fixed point
can be calculable from the FP equation at the boundary. 
Nevertheless the dual pair have the identical entropy .

It is also a very important clue that the Inflaton potential is concave
due to strong subadditivity of EE.
In fact, convex  potentials are observationally almost excluded by now.

The inflation is estimated to last                                                                                                                                                                                                                                                                                                                                                                                                                                                                                                                                                                                                                                                                                                                                                                                                                                                                                                                                                                                                                                                                                         $ 50 <N< 60$. 
On the other hand, $g\sim 10^{-10} $.
So we can regard the effect of $g$ to be tiny and safely negligible.
In the weak coupling regime correlated $g<\epsilon$, the theory becomes classical.

From holographic perspective, 
we have formulated the FP equation for the curvature perturbation at the boundary, i.e., the horizon of de Sitter space.

The von Neumann entropy of curvature perturbation  provides the quantum correction to the semiclassical dS entropy $S=1/g(t)$. 
 $g(t)$ undergoes time evolution due to the quantum effect,
i.e., a quantum inflation picture. 

We have investigated  two types of the solutions of GFP equation. 
The first category consists of the pre-inflation solutions (\ref{solution1}) which are UV complete .
The second category is the  inflationary space-time solution (\ref{solution2}) which is not UV complete ,
but the inflation terminates when $\epsilon$ grows up to $0(1)$. 
This transition scenario is consistent with the physical expectation that the solution with dominant entropy is chosen. 
Since the entropy $1/g$ for (\ref{solution1}) increases logarithmically 
and that for (\ref{solution2}) increases in a power-law, it is possible
the former is chosen initially and the latter is chosen if we wait long enough. 

The gravitational coupling may start with $g=1/2$ at the UV fixed point. 
The Universe may enter the inflation era. $50 < N <60, H/H_0 \sim 10^{-5}$ for linear potential.

Although it is an attractive scenario consistent with maximum entropy principle, it is another matter to reveal the transition 
process from pre-inflation to inflation theory.   

We need to clarify wether complete solutions exist or not. Another question is which combination of the solutions account for the Universe.
For UV completeness,  the pre-inflation solution (\ref{solution1}) is indispensable.
From the entropy considerations, the universe with small $n$ may be favored.
Although there are many interesting issues to be clarified, they are beyond the scope of this paper. 
Holography is a specific feature of quantum gravity.
In this paper we have derived the Fokker-Planck equation
on the boundary by using the technique of the renormalization group . We have proven the de Sitter duality between 
the bulk and the boundary located at the horizon. We find
de Sitter duality is deeply connected with the von Neumann / entangled entropy.
Suppose the total Hilbert space is factorized into
the bulk states and boundary states. The entropy of the bulk and
the boundary are the same. This explains why de Sitter duality exists at all.
It also explains why $\beta$ function at the UV fixed point
can be calculable from the FP equation at the boundary. 
Nevertheless the dual pair have the identical entropy .

It is also a very important clue that the Inflaton potential is concave
due to strong subadditivity of EE.
In fact, convex  potentials are observationally almost excluded by now.

The inflation is estimated to last                                                                                                                                                                                                                                                                                                                                                                                                                                                                                                                                                                                                                                                                                                                                                                                                                                                                                                                                                                                                                                                                                         $ 50 <N< 60$. 
On the other hand, $g\sim 10^{-10} $.
So we can regard the effect of $g$ to be tiny and safely negligible.
In the weak coupling regime $g<\epsilon$, the theory becomes classical.

Lastly, we stress the merit of Langevin formalism to  estimate $T / S , n_s$ and so on.
For example,
\begin{align}
S=4gN,~~
T =8g,~~
T/S=2/N .
\end{align}
$N$ is the total length of the random walk generated by curvature perturbation at the boundary.
They are 2 dimensional fractals. 
\section*{Acknowledgment}
This work is supported by Grant-in-Aid for Scientific Research (C) No. 16K05336. We thank Hiroyuki Kitamoto, 
Masashi Hazumi, Satoshi Iso, Hikaru Kawai, Kazunori Kohri, Takahiko Matsubara, Jun Nishimura, Hirotaka Sugawara, Takao Suyama and Yuko Urakawa for discussions.

\appendix

\section{Guide to de Sitter Space}
\setcounter{equation}{0}

Recently, de Sitter spacetime and inflation have drawn significant attention in the theoretical physics/superstring community. Some of the most interesting topics are holography and the thermodynamics associated with the de Sitter horizon. In this context, the static form of the metric of the de Sitter spacetime
is commonly used
 \begin{align}
ds^2 = - (1- H^2R^2)d\tau ^2 +\frac{dR^2}{1-H^2R^2}+R^2d\Omega^2
 \end{align}.
The classical result of \cite{Frolov2003}  is that observer at the origin  detects a thermal radiation from the de Sitter horizon.

The increase of the de Sitter entropy $S=1/g$ can be evaluated by the first law $T\Delta S=\Delta E$ 
where $\Delta E$ is the incoming energy flux of the inflaton  \cite{Frolov2003}.
Obviously we are estimating the entropy of sub-horizon.

We work in BRS type local gauge.

In our parametrization, the massless scalar field with the canonical kinetic term is
\begin{align}
X=2\sqrt{3}\omega-\frac{1}{\sqrt{3}}h^{00}.
\end{align}
We may focus on the  massless minimally coupled $X$ field to exlplore IR dynamics.

The curvature perturbation $\zeta$ 
is the fluctuation of the over all scale of the spacial metric:
 $g_{ij}=e^{2\zeta}\hat{g}_{ij}$.  
 Indeed we find $X$ reduces  to curvature perturbation $\zeta$
 as $X\sim \zeta$
 
 \begin{align}
X=&2\sqrt{3}\omega-\frac{1} {\sqrt{3}}h^{00}\\
=&(\sqrt{3}-\frac{1}{\sqrt{3}})h^{00}=\zeta
\end{align}

We find no IR logarithms in the BRS trivial sector.
The massless sector of Einstein gravity in de Sitter space consists of minimally coupled modes, 
and conformally coupled modes.   In the latter case, the coefficient
of the second de Witt-Schwinger expansion of the propagator cancels
as $R/6-2H^2=0$.
We can ignore the conformally coupled modes 
and focus on the massless minimally coupled modes .

The quadratic action is diagonalized as follows 
\begin{align}
\frac{1}{\kappa^2}\int d^4x\big[
&-\frac{1}{4}a_c^2\partial_\mu \tilde{h}^{ij}\partial^\mu \tilde{h}^{ij} 
+\frac{1}{2}a_c^2\partial_\mu X\partial^\mu X \notag\\
&+\frac{1}{2}a_c^2\partial_\mu h^{0i}\partial^\mu h^{0i}+H^2a_c^4h^{0i}h^{0i} 
-\frac{1}{2}a_c^2\partial_\mu Y\partial^\mu Y-H^2a_c^4 Y^2 \big], 
\label{diagonalized}\end{align}

where $X$ and $Y$ are given by 
\begin{align}
X\equiv 2\sqrt{3}\omega-\frac{1}{\sqrt{3}}h^{00},\hspace{1em}Y\equiv h^{00}-2\omega. 
\end{align}

Einstein gravity on de Sitter space receives  IR logarithmic quantum corrections.
We find the slow roll inflation theory is identical to Einstein gravity after taking account of this effect.
As a concrete example, we demonstrate the following single-field inflation model 
is dual to Einstein gravity with the identical IR logarithmic effects to the leading log order.

\begin{align}
\frac{1}{\kappa^2}\int d^4 x\sqrt{-g}\big[{R}
-6H^2(\gamma)\exp(-2\Gamma(\gamma )f) 
-2\Gamma(\gamma)g^{\mu\nu}\partial_\mu f \partial_\nu f\big]
\end{align}

It is clear from this Lagrangian that the inflaton $f$ rolls down an exponential potential. 
The Hubble parameter decreases as the Universe evolves and it eventually vanishes.

This action looks as follows if we make the conformal mode $a$ dependence explicit: 

\begin{align}
\frac{1}{\kappa^2}\int d^4 x\big[a^2\tilde{R} 
+6\tilde{g}^{\mu\nu}\partial_\mu a\partial_\nu a 
-6H^2(\gamma)a^4\exp(-2\Gamma(\gamma )f) 
-2\Gamma(\gamma)a^2\tilde{g}^{\mu\nu}\partial_\mu f \partial_\nu f\big], 
\label{Inflaton}\end{align}

We may eliminate the inflaton $f$ by using the following solution of (\ref{Inflaton}).
\begin{align}
a=e^f=a_c^{1+\gamma}. 
\end{align}

 The result deviates from the Einstein action as scaling dimensions of the cosmological constant is renormalized. 
 
\begin{align}
\frac{1}{\kappa^2}\int d^4 x\big[a^2\tilde{R} 
+(6-2\Gamma)\tilde{g}^{\mu\nu}\partial_\mu a \partial_\nu a 
-6H^2(\gamma)a^{4(1-\frac{\gamma}{2})} \big]. 
\label{CMact}\end{align}

The cosmological constant operator is made of the conformal mode $e^{4\omega}$.
We estimate the leading quantum renormalization.
\begin{align}
<e^{4\omega}>\sim 8<\omega^2>
\sim a^{-(3/2)g}
\end{align}
The anomalous dimension of cosmological constant operator is $a^{-2\gamma} 
=e^{-2\gamma Ht}$. We have finite renormalization of the kinetic term
of the conformal mode
$6 \rightarrow (6-2\Gamma) $.
In fact the same result is obtained by  quantizing  Einstein gravity
while inflaton is indispensable to provide counter terms
\cite{Kitamoto2019-2}.
Since $\epsilon= \gamma$, the inflation never terminates in this model.

\section{ Slow Roll Parameters}
\setcounter{equation}{0}

In Table 1.
\ref{tab:tm1}, we list  $\epsilon,1-n_s,dn_s/dN$ in the concave potential $V=N^{1/2n},n=1,2,3$. $n$ is a real parameter  but  we have chosen a plausible trio as $n=1$ has the largest entropy.
We note $n_s= \frac{1+1/2n}{\tilde{N}} $ is bounded from below by $1/\tilde{N}$ while
$8\epsilon =2/nN$ is not bounded from below when $n$ becomes large.
However $n=1$ is the maximum entropy concave state.
 It is smaller than $r=16\epsilon$ by a factor of 2.

\begin{align}
n_s=& \frac{\partial }{ \partial \tilde{N}} \log (4g) + \frac{\partial}{\partial N}\log(N)\\
  =&2\epsilon+\frac{1}{\tilde{N}}=\frac{1+1/2n}{\tilde{N}}
    \end{align}
  
  \begin{table}
\begin{center}
\begin{tabular}{ccccc}
&$r=2/nN$
& $n=1$
& $n=2$
&$n=3$
\\[.5pc] \hline\hline
&$\epsilon =1/4nN$
&$\frac {1}{4\tilde{N}}$
& $\frac {1}{8 \tilde{N}}$
& $\frac {1}{12\tilde{N}}$
\\[.5pc] \hline
&$n_s$
&$1-\frac{3}{2\tilde{N}}$
&$1-\frac{5}{4\tilde{N}}$
& $1-\frac{7}{6\tilde{N}}$
\\[.5pc] \hline\hline
&$dn_s/dN$
&$-\frac{3}{2\tilde{N}^2}$
&$-\frac{5}{4\tilde{N}^2}$
& $-\frac{7}{6\tilde{N}^2}$
\\[.5pc] \hline
\end{tabular}
\caption{\label{tab:tm1} }

\end{center}
\end{table}

\begin{table}
\begin{center}
\begin{tabular}{ccccc}
& $\tilde{N}=50$
& $n=1$
& $n=2$
&$n=3$
\\[.5pc] \hline
&$r =2/nN$
&$0.04$
& $0.02$
& $0.007$
\\[.5pc] \hline
&$n_s$
&$0.97$
&$0.975$
& $0.977$
\\[.5pc] \hline\hline
&$dn_s/dN$
&$-0.0006$
&$-0.0005$
& $-0.00046$
\\[.5pc] \hline
\end{tabular}
\caption{\label{tab:tm2} N=50}
\end{center}
\end{table}

Table II lists our numerical estimates by putting $\tilde{N}=60$ in Table I.

\begin{table}
\begin{center}
\begin{tabular}{ccccc}
& $\tilde{N}=60$
& $n=1$
& $n=2$
&$n=3$
\\[.5pc] \hline\hline
&$r $
&$0.033$
& $0.016$
& $0.008$
\\[.5pc] \hline
&$n_s$
&$0.975$
&$0.979$
& $0.981$
\\[.5pc] \hline\hline
&${dn_s/dN}$
&$-0.0006$
&$-0.0005$
& $-0.00046$
\\[.5pc] \hline
\end{tabular}
\caption{\label{tab:tm3},N=60}
\end{center}
\end{table}

  They are in the excellent agreement  with the current observations \cite{2018}.

   \begin{table}
\begin{center}
\begin{tabular}{cc}
&Planck 2018
\\[.5pc] \hline\hline
&$ r_{0.002}<0.066$
\\[.5pc] \hline
&$n_{s,0.002}=0.979 \pm 0.0041$
\\[.5pc] \hline
&$dn_s/dN = - 0.006 \pm 0.013$
\end{tabular}
\caption{\label{:tm4}, Planck 2018}
\end{center}
\end{table}
We adopt the same pivot scale $0.002Mpc^{-1}$ for $r,n_s$ to respect self-consistency.
It is important to establish the bound on $n$.
We recall here the curvature perturbation:
\begin{align}
P\sim 2.2 \times 10^{-9}.
\label{CP}
\end{align}
It is bounded from below $g>10^{-11}$ if $\epsilon>1/200$.

\section{ Entangled Entropy}
\setcounter{equation}{0}

Entangled entropy (EE) may be regarded as the Wilson loops (order parameters) in many body quantum systems.
The entangled entropy (EE) is a helpful bridge between gravity (string) and condensed matter physics. We concisely  list the important properties of EE here.
	See \cite{HolographicEE} for more details.

(1-1) Definition of EE:

Divide a quantum system into 2 parts $A,B$. 
\\The total Hilbert space becomes factorized.
\begin{align}
H_{tot}=&H_A\otimes H_B\\
\rho_A=&Tr_B \rho_{tot}\\
von ~~ Neumann~~entropy:&\nonumber\\
S_A&=-Tr_A\rho_A\log\rho_A
\end{align}

(1-2)  Basic Properties of EE :\\
If $\rho_{tot}$ is pure state, and $H_{tot}=H_A\otimes H_B$,
then $S_A=S_B$

(ii) Strong Subadditivity (SSA)  [Lieb-Ruskai 73] :\\
$S_{A+B}+S_{B+C}>
S_{A+B+C}+S_B$
\\
Inflaton potential is concave.

\end{document}